Original Paper

# Children with PIMD/SMID's expressive behaviors: Development and testing of ChildSIDE app, the first step for independent communication and mobility

Von Ralph Dane Marquez Herbuela[1], Tomonori Karita[1]*, Yoshiya Furukawa[2], Yoshinori Wada[1], Shuichiro Senba[3], Eiko Onishi[3], Tatsuo Saeki[3]

[1]Department of Special Needs Education, Graduate School of Education, Ehime University, Japan
[2]Graduate School of Humanities and Social Sciences, Hiroshima University, Japan
[3] DigitalPia Co., Ltd., Japan

*Corresponding Author:
Tomonori Karita, PhD
Department of Special Needs Education, Graduate School of Education, Ehime University, Bunkyo-cho 3, Matsuyama, 790-8577, Japan
Phone: +81-(0)89-927-9517
Email: karita.tomonori.mh@ehime-u.ac.jp

## Abstract

**Background**
Children with profound intellectual and multiple disabilities (PIMD) or severe motor and intellectual disabilities (SMID) only communicate through movements, vocalizations, body postures, muscle tensions, or facial expressions on a pre- or protosymbolic level. Yet, to the best of our knowledge, hardly any system has been developed to interpret their expressive behaviors. This paper describes the design, development, and testing of ChildSIDE in collecting children's behaviors and transmitting location and environmental data to the database. The movements associated with each behavior were also identified for future system development.

**Methods**
ChildSIDE app was pilot tested by conducting face-to-face and video-recorded sessions among purposively recruited child-caregiver dyads.

**Results**
ChildSIDE was more likely to collect more correct behavior data than the paper-based method (P = < .001) and it had >93% in detecting and transmitting location and environment data except for iBeacon data (82.3%). Behaviors were manifested mainly through hand (22.8%) and body movements (27.7%), and vocalizations (21.6%).

**Conclusions**
ChildSIDE is an effective method in collecting children's expressive behaviors with a high accuracy rate in detecting and transmitting environment and outdoor location data. There is a need for a system that uses motion capture and trajectory analyses for developing algorithms to predict children's needs.

**Keywords:** profound intellectual and multiple disabilities, severe motor and intellectual disabilities, mobile app development, AAC, smartphone-based data collection

# 1. Introduction

Children with profound intellectual and multiple disabilities (PIMD) or severe motor and intellectual disabilities (SMID), as the name implies, have an estimated intelligence quotient of less than 25 (IQ < 25) which is equivalent to a maximum developmental age of 24 months [1,2]. They often have difficulty in communication especially understanding spoken or verbal language and symbolic interaction with objects [3,1]. They also have severe or profound motor disabilities characterized by restricted or absence of hand, arm, and leg functions that cause limited or lack of ability to move by themselves [1,4]. Sometimes, these children have sensory impairments and chronic health conditions which include but not limited to epilepsy, visual impairments, constipation, spasticity, deformations, incontinence, and reflux [5,6]. Despite the severe challenges brought by their condition, these children should, consequently, be able to communicate with people and interact with the environment independently.

Due to their profound intellectual and neuromotor disabilities, one of the most challenging parts of supporting these children is communication. Several augmented alternative communication (AAC) applications (apps) have been developed that focus on helping children with speech disabilities and one of which is voice output communication aid or VOCA. With the help of mobile phones, VOCA apps like Drop Talk and Voice4U have been helping children with speech disabilities communicate with other people. Their main function is to produce a voice when a user clicks a specific icon, symbol, or picture (display) that corresponds to a word or phrase. These displays can be combined (interface) to make sentences in order to match a specific situation. While this is a promising support approach for children with speech disabilities, selecting displays and choosing interfaces that best fit a specific situation is quite difficult for children with speech and/or intellectual disabilities (ID), because of their inability to determine which interface they should switch to in each situation and location due to their cognitive disability [1].

We have developed Friendly VOCA, a user-friendly VOCA iOS mobile app that enables children and individuals with speech and/or IDs to communicate with other people independently [7]. Unlike other available VOCAs, Friendly VOCA has the ability to automatically switch displays or interfaces that match the user's location at a specific time [7]. It uses Global Positioning System (GPS) to identify the user's current outdoor location in terms of map coordinates (latitude and longitude). However, since GPS has a limited ability to identify indoor locations (e.g. inside a store or a room) and elevated surfaces (building floors, etc.), we used iBeacon. It is a system developed by Apple Inc. that is based on Bluetooth low energy proximity sensing which transmits a universally unique identifier (UUID) to a user's app. These two combined systems have helped Friendly VOCA to switch interfaces and displays automatically depending on the user's location at a specific time. Both GPS and iBeacon systems have been tested and experiments revealed that they can automatically show appropriate interfaces and displays that correspond to user's locations with 100% and 71% accuracy, respectively [7].

Grounded in Schank and Abelson's Script Theory [8], Friendly VOCA's concept of automatically switching displays or interfaces that match the user's location is based on the notion of "scripts." Scripts are the organized set or body of our basic background knowledge or "schema" that we must have in order to understand how we respond or behave appropriately to a particular situation or location [8]. This is best explained in Schank's classic example of scripts in a restaurant: when we enter a restaurant, we greet the waiter or we look for a table where to sit, we approach the table, we sit in a sitting position, then we browse the menu, we decide what to order, then we call the waiter and the waiter comes to our table and we say what food we want to eat, then the waiter tells the chef, then the chef cooks our food, then the chef gives the food to the waiter and the waiter brings it to us, then we eat the food, after eating, the waiter writes the check, gives it to us, we give money to the waiter and we go out of the restaurant [8]. We have demonstrated this theory in Friendly VOCA through the use of our schema on the specific scripts in the form of varied displays and interfaces tailored to a specific situation (e.g. class or playtime), location (e.g. classroom, playground, home), and time (e.g. morning, lunch breaks, evening) using GPS and iBeacon systems.

While the use of scripts greatly matches Friendly VOCA, it may also present possible misunderstandings or incorrect inferences due to many variations of the situations or locations (e.g. type of restaurant) where a general script may not be applicable (e.g. different scripts in fast-food and fine dine-in). Similarly, Friendly

VOCA's set of displays and interfaces may not perfectly cater to all the children with speech and/or IDs since each of them has their own, individual needs that are beyond what Friendly VOCA can provide. Most importantly, it also leaves the children with PIMD/SMID, behind. Since it requires the user to choose and click an icon or symbol to produce a voice output, it needs an apparent understanding of symbolic interaction (interpreting symbols or icons) or verbal language (comprehending voice outputs) which may seem difficult for these children due to their severe or profound IDs [1]. These children may not understand that a symbol or a picture that shows a hand with its index finger pointing to a face means "I," "I am," or "me," more so, the meaning of the voice output that corresponds to what it means. Moreover, clicking an icon or symbol can also be physically demanding for them because of their profound neuromotor dysfunctions [1].

Children with PIMD/SMID only communicate through movements, sounds, body postures, muscle tensions, or facial expressions on a presymbolic (nonsymbolic) or protosymbolic (limited information) level with no shared meaning, which hinders expressing their needs [9,10,11,12]. These behaviors can also be minute and refined which may be difficult for caregivers and teachers to perceive and interpret their needs [9]. Surprisingly, to our knowledge, before Tanaka (2000) [13], Motoda (2002) [14] and Ashida and Ishikura [9,15], scarcely any study had examined the behaviors of these children to enable perception and interpretation. In 2013, Ashida and Ishikura [9] introduced six major categories based on the body parts movements involved in each expressive behavior of children with PIMD/SMID: eye movement, facial expression, vocalization, hand movement, body posture, body movement and non-communicative behaviors (others). They then used these categories in analyzing the expressive behaviors of two children in 2015 [15]. They found out that one child had many active movements of arms, legs, and eyes, and expressed needs and emotions by changing gaze and smiling, while the expressions of the other child were limited to the movements of the head, neck, mouth, and eyes [15]. This suggests that in order to predict the needs of children with PIMD/SMID, interventions that focus on interpreting their expressive behaviors, whether it involves the head, face, or upper limb movements, can be developed. However, in order to do this, first, we have to collect children's expressive behaviors associated with their needs. Yet, to the best of our knowledge, there is hardly any technology developed for this purpose.

Although the paper-based method can be used in collecting children's behavior data because it is more convenient to use, it is also more prone to a higher frequency of incomplete records and potential human errors than the use of smartphone-based data collection which provides real-time data and is more efficient and accurate with minimal errors and inconsistencies [16,17]. Thus, we developed ChildSIDE, a mobile app that collects children with PIMD/SMID's expressive behaviors as interpreted by their caregivers. Furthermore, since apps can combine data from other smartphone-based features like GPS [17], similar to Friendly VOCA and grounded from the notion of scripts, the ChildSIDE app also collects location and environment data through the use of sensing technologies. By not only collecting and analyzing children's behaviors but also by collecting and analyzing location and environment data that are associated with each behavior, we will be able to infer their intentions and needs in the future.

This paper describes the design and development of the ChildSIDE app. We pilot tested it among purposively recruited children with PIMD/SMID and their caregivers and evaluated its accuracy in terms of collecting their expressive behaviors and the time each behavior occurred (timestamps) compared with paper-based data collection methods, and its accuracy in terms of detecting and transmitting location and environment data to the app database. We also sought to identify which movements are associated with their expressive behaviors by categorizing them using the category table of expressions by Ashida and Ishikura (2013) [9]. This will help in identifying the method or design of the system that we will develop in the future. Based on previous literature, we hypothesized that the ChildSIDE app, compared with the paper-based collection method, will more likely to accurately record correct behavior data with minimal incorrect or missing data brought about by human errors. This study is exploratory in the context of testing the app's ability to accurately detecting and transmitting environmental data using the sensors and API, but not the use of GPS and iBeacon systems. Due to the relatively low accuracy rate of iBeacon based on our previous experiment, we decided to use another brand of iBeacon for this study. Thus, we hypothesized that the ChildSIDE app would yield higher accuracy rates in detecting and transmitting indoor location data to the app database. Lastly, we hypothesized that children's behavior will mostly involve upper limbs and body movements similar to the reports of the previous study [9].

## 2. Methods
### 2.1. ChildSIDE App Design and Development

We have developed ChildSIDE, a mobile app which collects: a) caregivers' interpretation of children with PIMD/SMID's expressive behaviors and timestamps; b) location, and; c) environment data (Figure 1). It was developed in Android (7.0) mobile platform (HUAWEI P9 lite) using Eclipse Android Studio (version 4.0.1), an integrated development programming environment software and Java 1.80_242 (OPEN JDK) programming language. ChildSIDE's design was based on the IoT (internet of things) systems for human interaction interface (HCI) and its name originated from its main goal of being "beside" of its target population and children (Child) by aiding independent communication and mobility. SIDE also stands for "Sampling Information and Data of children's expressive behaviors and the Environment" which is explicitly derived from its main function of collecting children's expressive behaviors with associated environmental data.

ChildSIDE has two interfaces: Behavior settings interface (**Figure 1.a**.) and the Behavior list interface (**Figure 1.b**.). Behavior settings interface allows the user to add behavior by putting category code, the behavior's name, and the category name. When adding new behavior, users need to click the "Add row" button, then a new row will appear in the list above it. Users should click that blank row to enter the category code (assigned codes for the order of behaviors where most common behavior is coded "0," and it appears on top of the list and the second most common behavior is coded "1," which follows the behavior that was coded 1, etc.), the expressive behavior's name, and a category name in the settings interface below it. To save it, users should click the "update" button, then the new behavior with its corresponding code and category name will appear in the list above the setting interface and the Behavior list. When adding a new behavior to an already existing category, users need to enter the name of the category, otherwise, a new category will be created. Categorizing the behaviors will make it easier for the user to locate or update them.

**Figure 2** shows how the data were detected by the ChildSIDE app from the data sources (iBeacon, GPS, ALPS Sensors, and OpenWeatherMap API) and transmitted to the Google Firebase database. We used Android's built-in time stamps and GPS (a) to identify the subject's current outdoor location in terms of map coordinates (latitude and longitude). We also used iBeacon (b) (Braveridge BVMCN1101AA B), which transmits UUID, Radio Signal Strength Indication (RSSI), and iBeacon name to the nearest app to identify the subject's specific indoor location. If there are multiple iBeacons installed in a location, the app detects the UUID of the nearest iBeacon. We also used a multi-function ALPS Bluetooth sensors (c) module (688-UGWZ3AA001A Sensor Network Kit W/BLE Mod Sensors) to acquire and transmit 11 motion and environmental data: temperature and humidity (℃ and %RH); geomagnetic sensor (electric compass; 6-axis Accel+Geomag) (ranges: g1, g2, g3 and resolutions: µT1, µT2, µT3); ultraviolet (UV) or ambient light (mW/cm2 and Lx), and; atmospheric pressure (hPa). Weather information (weather, pressure, humidity, sunrise, and sunset) was obtained from OpenWeatherMap Application Programming Interface (API) (d), an online service that provides weather data that matches the user's current location. It has 15 parameters: country name, location name (region or city), weather, sunset time, sunrise time, current time, minimum temperature (℃), maximum temperature (℃), atmospheric pressure (hPa), main temperature (℃), humidity (%), weather description, cloudiness (%), wind direction (degrees), and wind speed (meter/sec.). When a user clicks a behavior, the app automatically sends the behavior and category name with its associated GPS and iBeacon location data, and environment data from the OpenWeatherMap API and the ALPS sensors to the Google Firebase database (e), a third-party service provider that allows the data to be stored in real-time and synchronized among mobile platforms.

### 2.2. Study and sampling designs and participant inclusion criteria

This pilot testing utilized a cross-sectional-observational study design using multiple single-subject face-to-face and video-recorded sessions. Studies that used single-subject design among children with special education needs showed more powerful results than those studies that used a group research design [18]. This study involved purposively sampled child-caregiver dyads recruited at a special needs school from September 2019 to February 2020. Children included in this study met these following criteria: a. diagnosed with PIMD/SMID, or; b. severe or profound ID; c. with or without comorbid sensory impairments and/or chronic health conditions which include but not limited to epilepsy, visual impairments, constipation, spasticity, deformations, incontinence, and reflux, etc. d. whose chronological or mental age were 18 years and below at the time of the study. Caregivers can either be the primary caregivers (immediate family members) or secondary caregivers (non-family like teachers, supporters, etc.) who have been living or supporting the children for three years or more. This criterion was set to ensure that caregivers were familiar and have a schema about the children's expressive behaviors.

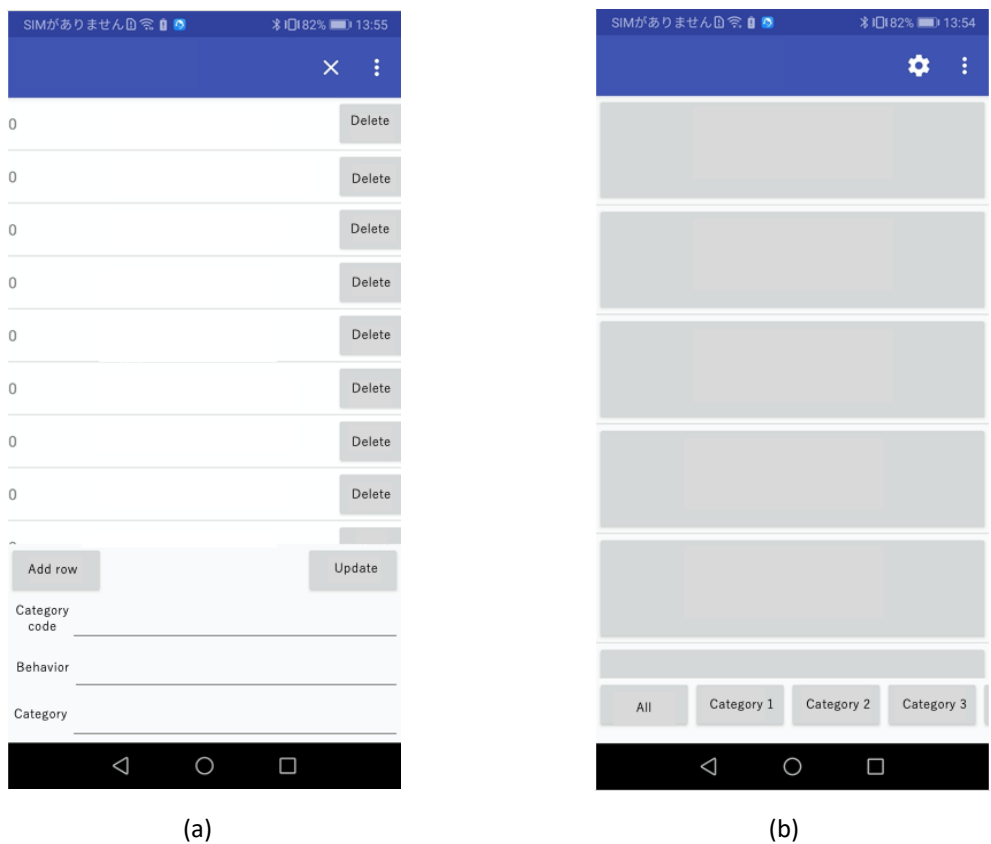

(a) (b)

Figure 1. ChildSIDE User Interface. (a) behavior settings, (b) behavior list.

### 2.3. Ethical Considerations

This study was written and conducted per the Declaration of Helsinki [19] and the International Council for Harmonization Good Clinical Practice guidelines [20]. This is part of a project that was approved by the Ehime University, Faculty of Education Research Ethics Committee (approval number: R2-18). The primary caregivers of all the participants gave their consent for their child's participation in this study by signing a written informed consent. They were also informed that their child's participation in the study was voluntary and they may stop their participation at any time. All data that contain participant information or identity were coded (video recordings were blurred) and stored in a password-protected database and computer for their protection and privacy.

### 2.4. Intervention

We used our previously designed and tested intervention setup (Figure 3) with multiple single-subject face-to-face sessions conducted in a classroom setting. The duration of each session depends on the child's availability and the willingness of their caregivers to participate and be recorded. We used a video-based recording method in all the sessions for inter-rater analyses and categorizing the behaviors. This method has been reported to have higher inter-rater reliability (than traditional observational methods) and allows researchers to collect, analyze, and validate data retrospectively [21]. We used one videotape recorder (VTR) in a tripod which was placed two meters from the participants to capture the child's facial expressions and upper and lower limb movements (**Figure 3.a.**), and all the exchanges of responses between the child and their caregiver (**Figure 3.b.**). Before the intervention, we installed one iBeacon device in each of the eight classrooms where we held the sessions. We placed one iBeacon and one ALPS sensor on a table near the investigator 1. Investigator 1 recorded the children's expressive behaviors that were associated with their needs in the ChildSIDE app while Investigator 2 used a paper-based collection method. All sessions were recorded in the classrooms where the children usually spend time so they can behave normally and interact with their caregivers even with the presence of other children and caregivers. For this same reason, the sessions targeted morning greetings, lunchtime, and break time. When a child shows some reaction (e.g. vocalization, gesture), the caregiver responds by confirming the child's need (e.g. want to go to the toilet) verbally or by actions (e.g. assist the child to the toilet).

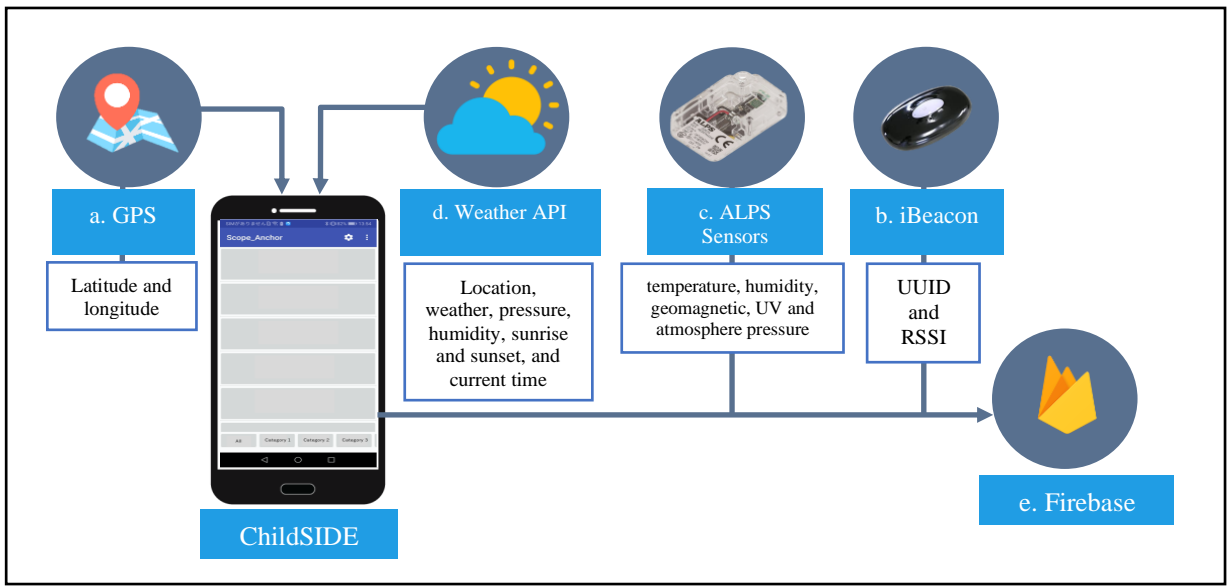

**Figure 2.** Data flow from the data sources (iBeacon, GPS, ALPS Sensors and OpenWeatherMap API) detected and transmitted by ChildSIDE app to Google Firebase database.

**2.5. Statistical and Data Analysis**

We compared the accuracy of the app and the paper-based collection method in collecting behavior data time stamps. One investigator who has expertise on behaviors and support of children with PIMD/SMID watched the video recordings and made a separate list of the behaviors and time stamps. This served as our reference database which we used to compare the data transmitted to the app database and the data on the paper. Each behavior and time stamp in the app database and on the paper was labeled "correct" and was scored "1" if it matches the one in the reference database while missing (not in the reference database either by an app system error or human error such as forgetting to write or click the button in the app) and incorrect (did not match the ones in the reference database due to app system error or human error such as clicking the wrong button) data were scored "0". After deleting all the test data in the app database, we computed for the chi-square test of association between the percentage of correct and missing or incorrect data of ChildSIDE and the paper-based method. Odds ratio effect sizes were also computed to measure the differences in the proportion of correct and missing or incorrect data between the app and the paper-based collection method.

To measure the accuracy of ChildSIDE in detecting and sending data from data sources (iBeacon, GPS, ALPS sensors and OpenWeatherMap API) to the database, we computed for the frequency distribution and percentages (%) of the 31 location, motion, and environment data types from each data source: iBeacon (3), GPS (2), ALPS sensors (11), and OpenWeatherMap API (15). We deleted those behavior data without any associated data from any data source. Each data transmitted to the app database was scored "1" while errors (app fails to detect signals from sensors or vice versa) were scored "0". Since each data source has multiple data types, we computed for the mean scores and compared them with the total number of behaviors data.

To classify which body parts or movements (minor categories) were involved in each behavior using the table of expressions in children with PIMD/SMID (**Table 1**) (Ashida and Ishikura, 2013) [9], two raters watched the video recordings independently and analyzed each behavior recorded by the app. Each rater scored "1" in each minor category where a behavior belongs to, otherwise, they scored it "0". For example, "Goodbye" can be shown by waving the hands and producing sound, thus, this behavior will be given one score for moving (d.3.) minor category under hand movement and one score for vocalization (c) major category. To identify the kappa coefficients in each major category, we scored "1" in each major category where it had at least one minor category with a score of at least one. To test the agreement between the two raters in each behavior per minor and major category, we computed for Kappa statistics. Kappa ranges (0 = less than chance; 1.01-0.20 = slight; 0.21-0.40 = fair; 0.41-0.60 = moderate; 0.61-0.80 = substantial; 0.81-0.99 = almost perfect) were used to identify the level of agreement between the two raters in each major and minor category using the Kappa coefficients with a significance level of less than .01 p-value [22]. The two raters also counted the number of times (frequency) each movement (minor category) was shown in each behavior. Lastly, they reanalyzed their responses and once a consensus was reached, a final categorization of behaviors was created based on the table of expressions in children with PIMD/SMID. Chi-square and kappa statistical analyses were conducted using the 'stats' (version 4.0.1) and 'irr' (version 0.84.1) packages of R (version 4.0.2) statistical computing software.

## 3. Results
### 3.1. Participants profile

We were able to recruit 19 out of 22 child-caregiver dyads (three were excluded for unavailability) who were assessed for eligibility. Children had ages from eight to 16 years old (3rd grade to 1st year high school), 13 (68%) were males, and 15 (79%) had PIMD/SMID diagnoses while 4 (21%) had severe or profound intellectual disabilities. In total, we were able to conduct 105 sessions that ranged from as low as one session and as many as 15 sessions per child, and with an average of five sessions per child. **Figure 4** shows the participant, session, and data flow using the CONSORT (Consolidated Standards of Reporting Trials) diagram.

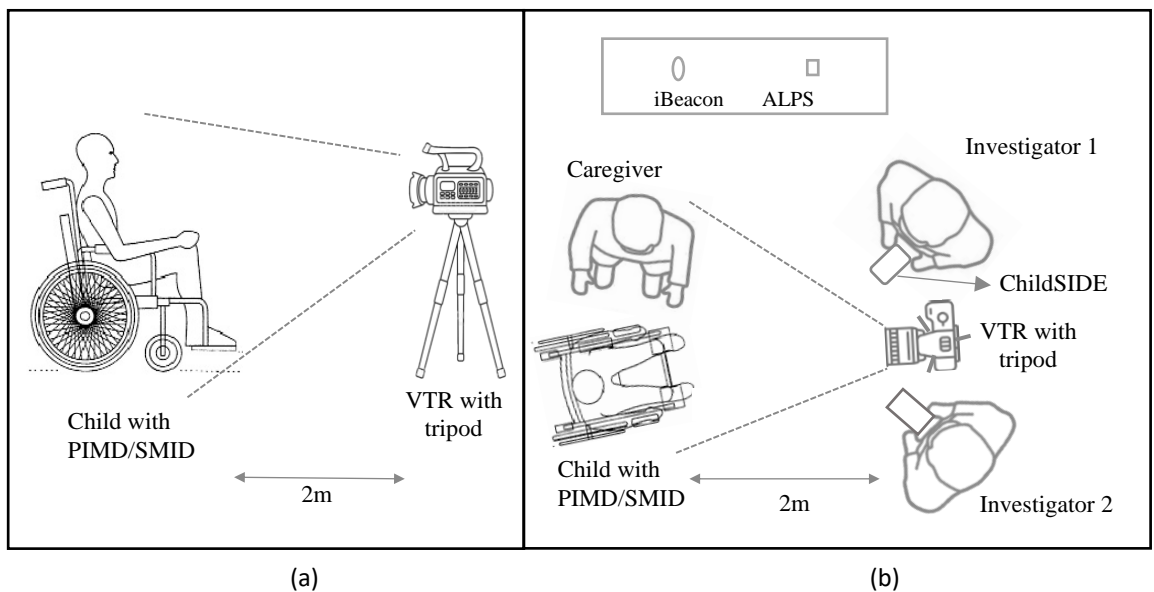

**Figure 3.** Intervention set-up. (a) Videotape recorder (VTR) focus on the facial, upper and lower limbs movements, (b) Intervention setup in a classroom setting: 2-meter distance from the VTR to the child with PIMD/SMID and caregiver, and the location where the iBeacon and ALPS sensors were placed.

**3.2. ChildSIDE vs paper-based data collection**

From the 90 sessions (ranged from 0.37 to 32 minutes video recording time, with a mean of 19 minutes and a standard deviation [SD] of 11.3) that we conducted, we were able to collect 308 individual behavior data. Of which, seven which were found to be test data were excluded bringing the total number to 301. Chi-square test demonstrated that the ChildSIDE app had significantly (P = < .001) more correct data (n = 269; 89.4%) than the paper-based collection method (n = 195; 64.8%) ($X^2$ = 51.48, df = 1). This represents the fact that based on the odds ratio, the ChildSIDE app was 4.6 times more likely to collect correct data and 0.2 times more likely to contain missing or incorrect data than the paper-based collection method.

**Table. 1.** Category table of expressions in children with PIMD/SMID [9]

| Categories | Criteria |
| --- | --- |
| **a. Eye movement** | |
| 1. Gazing | Gaze at people and things (in the case of interpersonal people, look at their faces) |
| 2. Eye tracking | Eye movements that follow the movements of people and things in a linear fashion |
| 3. Changing line of sight | Change of line of sight, movement of line of sight; gaze rolls and moves; point-like movement that is not "a.2. eye tracking." The momentary glare can also be evaluated. Movements that cannot be evaluated as gaze/tracking. |
| 4. Opening or closing the eyelids | Not an involuntary blink. Their reaction when told to open or close their eyes. |
| **b. Facial Expression** | |
| 1. Smiling | Smile |
| 2. Facial expression (other than smile) | Something that is not expressionless. Changes in facial expressions. Surprise, frowning, sticking out tongue, etc. |
| 3. Concentrating and listening | Focusing on picture books, music, and voices etc. |
| **c. Vocalization** | Producing sound |
| **d. Hand movement** | |
| 1. Pointing | Hand pointing or pointing finger towards an object. |
| 2. Reaching | The action of reaching or chasing after reaching the target, not by pointing hand or finger. |
| 3. Moving | Grab, hit, beckon, push, raise hands, dispel, etc. |
| **e. Body movement** | |
| 1. Approaching | Head or upper body (or the whole body) is brought close to a person or an object. |
| 2. Contacting | Touching people and things with hands and body. It does not include cases that are touched by accident or touched. |
| 3. Movement of a part of the body | Head and neck movements, upper body movements, upper and lower limb movements (shake, bend, move mouth, flutter legs, etc.); (excluding "d.1. pointing", "d.2. reaching", "d.3. moving"), etc. Distinguish from "f.1. stereotyped behavior" |
| **f. Non-Communicative Behaviors (Others)** | |
| 1. Stereotypical behavior | The same behavior or movement are repeated without purpose. Behavior that occurs in a certain repetition e.g. Finger sucking, shaking hands, rocking, etc. (Shaking things is "d.3. moving") |
| 2. Self- and others-injurious behavior | Hitting someone, biting finger, etc. |
| 3. Others | Difficult to classify other than the above categories |

### 3.3. iBeacon, GPS, ALPS Sensor and OpenWeatherMap API data

From the 19 child-caregiver dyads, one child-caregiver dyad (8-year-old male with PIMD/SMID), 15 additional sessions (ranged from six to 54 minutes video recording time [M = 28 minutes; SD = 13.8]) and 63 individual behavior data were added, bringing the total to 20 child-caregiver dyads, 150 sessions, and 364 individual behavior data, respectively. Of the 371 collected individual behavior data, there were 327 that had associated data from iBeacon, GPS, ALPS

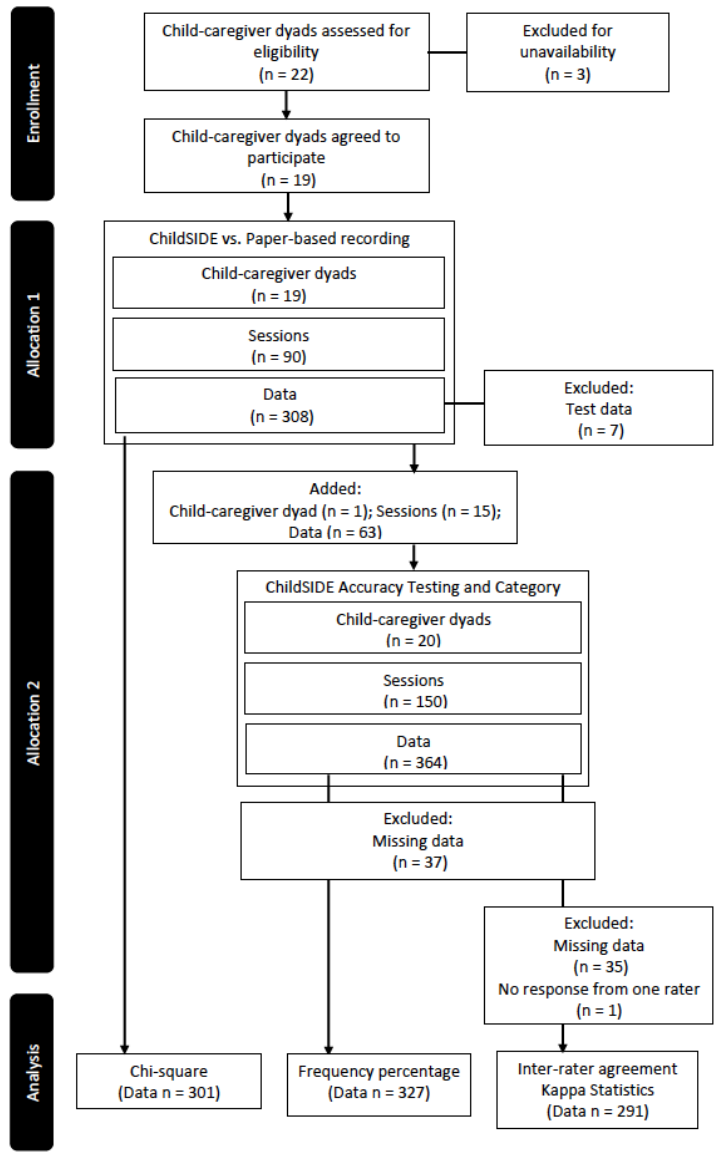

**Figure 4.** CONSORT diagram of participant, session and data flow from enrollment, allocation and analysis.

Sensor, or OpenWeatherMap API data sources. In addition to the seven previously deleted test data, we also deleted 37 needs data without any associated data from any data source. The app was able to detect and transmit 269 (82.3%) UUID, 269 (82.3%) RSSI, and 269 (82.3%) iBeacon names data (**Figure 5**). UV or ambient light sensors range (mW/cm2) (S1) and resolution (Lx) (S8) had the relatively lowest scores of 213 (65.1%) and 266 (81.3 %), respectively, among the ALPS sensors. 6-axis (Accel+Geomag) sensor ranges [g] (S2, S3, S4) and resolutions (μT) (S5, S6, S7) had score ranged from 318 (97.2%) to 321 (98.2%). Among the ALPS sensors, 100% of pressure sensor range (hPa) (S9), temperature and humidity sensor range (℃) (S10), and resolution (%RH) (S11) data were detected and transmitted by the app to the database. Among the OpenWeatherMap API parameters, wind direction (A14) had a relatively lower score of 288 (88.1%) compared with other parameters

(A1 to A13 and A15) that had scores of 312 (95.4%). In general, iBeacon had the relatively lowest mean score (M = 269; 82.3%) among the data sources: GPS (M = 327; 100%), ALPS Sensors (M = 305; 93.4%) and OpenWeatherMap API (M = 310; 94.9%). This means that the ChildSIDE app has an accuracy level that ranged from 82% to 100% in detecting and transmitting location and environment data to the database.

### 3.4. Expressive behavior category

In addition to the previously deleted test data (n = 7), needs data without any associated data from any data source (n = 37), we also deleted 35 individual behavior data that were not detected by the app. We were able to collect 292 individual behavior data (please see Table S1 for more information on the individual behavior data), of which, one individual behavior data had no score from the two raters, subjecting only the remaining 291 to inter-agreement Kappa statistics analysis. Table 2 shows the levels of agreement based on the Kappa coefficients and range, between the two raters in identifying the body parts or movements (minor categories) involved in each behavior. Kappa statistics revealed that the levels of agreement between the two raters in 14 out of 16 minor categories based on Kappa coefficients, ranged from fair (0.21-0.40) to almost perfect (0.81-0.99) with significant p-values of < .001. The minor categories which had the highest and lowest kappa coefficients were pointing and stereotypical behaviors with kappa coefficients of 0.88 and 0.21, respectively. Only one rater scored a need under the concentrating and listening category and the behaviors that fell under the self- and others-injurious behavior category were not the same between the two raters. Further, while the two raters had an almost perfect level of agreement (with significant p-values of < .001) in vocalization (0.95), hand movement (0.88) and substantial level of agreement in eye movement (0.83), facial expression (0.70), and body movement (0.78), non-communicative behaviors (Others) only had a kappa coefficient of 0.40 with a fair inter-rater agreement level. From these results, we were able to identify 676 body parts or movements involved in 291 individual behavior data. Of the 676, children's behaviors were composed of 27.7% body movement, 22.8% hand movement, 21.6% vocalization, 15.4% eye movement, 9% facial expression, and 3.6% other expressions.

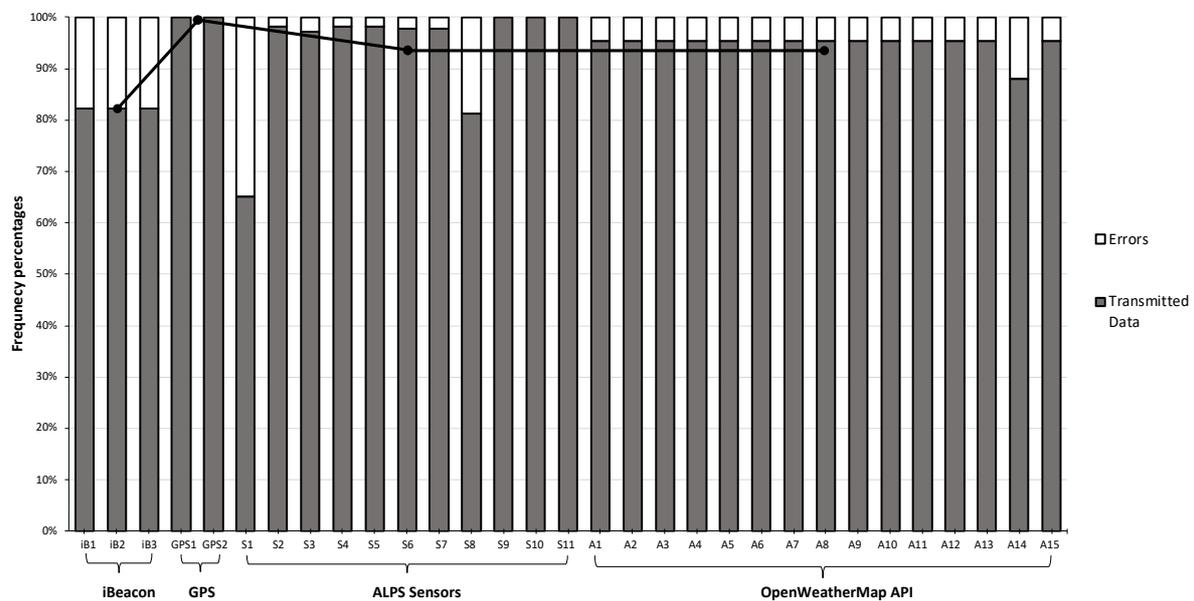

**Figure 5.** Frequency percentages of transmitted data and errors in each data type and mean scores (black dots) in each data source detected and transmitted by the ChildSIDE app to the database. iB1 = universally unique identifier (UUID); iB2 = Radio Signal Strength Indication (RSSI); iB3 = iBeacon name; GPS1 = Longitude; GPS2 = Latitude; S1 = UV range [mW/cm2]; S2, S3, S4 = 6-axis (Accel+Geomag) sensor ranges [g]; S5, S6, S7 = 6-axis (Accel+Geomag) sensor resolutions [µT]; S8 = UV resolution [Lx]; S9 = pressure sensor range [hPa]; S10 = temperature and humidity sensor range [°C]; S11 = temperature and humidity sensor resolution [%RH]; A1 =

country name; A2 = location name (region or city); A3 = weather; A4 = sunset time; A5 = sunrise time; A6 = current time; A7 = minimum temperature (℃); A8 = maximum temperature (℃); A9 = atmospheric pressure (hPa); A10 = main temperature (℃); A11 = humidity (%); A12 = weather description; A13 = cloudiness (%); A14 = wind direction (degrees); A15 = wind speed (meter/sec.).

Table 2. Inter-rater agreement Kappa statistics and frequency distribution and percentages of the major and minor categories of the table of expressions in children with PIMD/SMID [9]

| Categories | Inter-rater agreement | | | Frequency Distribution and Percentages |
|---|---|---|---|---|
| | Kappa coefficients | Kappa range | p-value | n = 676 (%) |
| **a. Eye movement** | 0.83 | 5 | < .001 | 104 (15.4) |
| 1. Gazing | 0.64 | 4 | < .001 | 38 (5.6) |
| 2. Eye tracking | 0.50 | 3 | < .001 | 13 (1.9) |
| 3. Changing line of sight | 0.53 | 3 | < .001 | 46 (6.8) |
| 4. Opening or closing the eyelids | 0.74 | 4 | < .001 | 7 (1.0) |
| **b. Facial Expression** | 0.70 | 4 | < .001 | 61 (9.0) |
| 1. Smiling | 0.69 | 4 | < .001 | 36 (5.3) |
| 2. Facial expression (other than smile) | 0.34 | 2 | < .001 | 24 (3.6) |
| 3. Concentrating and listening[a] | - | - | - | 1 (0.1) |
| **c. Vocalization** | 0.95 | 5 | < .001 | 146 (21.6) |
| **d. Hand movement** | 0.88 | 5 | < .001 | 154 (22.8) |
| 1. Pointing | 0.88 | 5 | < .001 | 29 (4.3) |
| 2. Reaching | 0.69 | 4 | < .001 | 25 (3.7) |
| 3. Moving | 0.79 | 4 | < .001 | 100 (14.8) |
| **e. Body movement** | 0.78 | 4 | < .001 | 187 (27.7) |
| 1. Approaching | 0.44 | 3 | < .001 | 16 (2.4) |
| 2. Contacting | 0.76 | 2 | < .001 | 35 (5.2) |
| 3. Movement of a part of the body | 0.64 | 2 | < .001 | 136 (20.1) |
| **f. Non-Communicative Behaviors (Others)** | 0.40 | 3 | < .001 | 24 (3.6) |
| 1. Stereotypical behavior | 0.21 | 2 | < .001 | 16 (2.4) |
| 2. Self- and others-injurious behavior[b] | -0.0003 | - | 0.95 | 2 (0.3) |
| 3. Others | 0.44 | 4 | < 0.001 | 6 (0.9) |

a = one score from one rater; b = needs did not match; Kappa ranges: 0 = less than chance; 1 = 1.01-0.20; 2 = 0.21-0.40; 3 = 0.41-0.60; 4 = 0.61-0.80; 5 = 0.81-0.99 [22]

## 4. Discussion

With the use of location and environmental sensing technologies, we were able to develop ChildSIDE, a mobile app that collects caregivers' interpretation of children with PIMD/SMID's expressive behaviors, location, and environment data. The app had significantly (P = < .001) more likely to have correct records and less likely to have missing or incorrect data than the paper-based collection method. The app was also able to detect and transmit data to the app with above 93% accuracy except for iBeacon which had the relatively lowest accuracy rate of 82.3%. Further, by conducting inter-rater Kappa statistics analysis which shown an almost perfect level of agreement between two raters, we were able to identify and categorize 676 body parts or movements involved in 291 individual behavior data, and we found out that expressive behaviors of children with PIMD/SMID were manifested mainly through body and hand movements and vocalizations.

The result that the app was more likely to have correct records and less likely to have missing or incorrect data than the paper-based collection method is in line with our hypothesis. Previous studies that investigated the difference between the two data collection methods, also found out that the smart-phone collection method provides timely data with fewer errors and inconsistencies than the paper-based collection methods [18,19]. When it comes to time, we spent almost two weeks in cleaning the paper-based data ready for encoding. Encoding to the analysis of paper collected data also took us approximately one week, compared to the data collected by the app which was readily available for analysis.

Among the location and environment data sensing technologies that we used, the app had a relatively lowest accuracy rate detecting and transmitting iBeacon data. Although relatively higher, our previous experiment on the use of the iBeacon system in Friendly VOCA showed the same results. This trend emphasizes the possible problem with the placement of iBeacon devices and not the mobile apps that we developed. That is, our intervention setup may be problematic since we put an iBeacon device approximately two meters from the app. Dalkilic et al (2017) [23] tested the accuracy of iBeacon devices in sending signals to an app, and they found out that when iBeacon is close to a mobile phone, the app has difficulty in detecting exactly where the signal is coming from. According to them, the electromagnetic fields or waves generated by mobile phones interfere with the ones coming from the iBeacon device, thus, low location accuracy. Their experiments also revealed that when iBeacon devices are placed away from mobile phones (if there are no radio interference from other iBeacon devices, laptops, or mobile phones), up to eight meters, the app gives more accurate distance estimations. Aside from this, we also thought that putting iBeacon devices in adjacent rooms caused the difficulty for the app to detect and therefore transmit iBeacon data to the app database. Thus, we checked if the iBeacon data detected and transmitted by the app to the database was from the iBeacon installed in the same room. We found out that the iBeacon data detected and transmitted by the app to the database were approximately 96% the same as the iBeacon installed in the same room as the app. This finding is similar to that of Dalkilic et al (2017) [23]. They examined the effect of walls by putting one iBeacon device and a mobile phone in one room and putting another iBeacon in an adjacent room. They found out that the wall between the two rooms blocked the signals from the iBeacon which was not in the same room as the app.

While we acknowledge and plan to address the problems in our intervention set up specifically with the placement of iBeacon devices when it comes to its distance with the app, we also assumed that the problem was caused by the signal strength of the iBeacon device that we used although it was different from the one we used before. Paek, Ko, and Shin (2016) [24] tested three iBeacon devices and they found out that the variation in signal was too high and the RSSI values and the corresponding signal propagation model vary significantly across iBeacon vendors and mobile platforms. To address this, we plan to test different iBeacon devices from different vendors and choose the best product that fits our mobile platform and the goal of our study in the future. Most importantly, we will also consider an iBeacon device company that conforms to the regulations and technical standards of Japan Radio Wave Law [7].

One of the main strengths of this study was the inclusion of a relatively higher number of children with PIMD/SMID or severe or profound IDs (n = 20) than similar previous studies which only had a maximum of two children. This enabled us to conduct 105 multiple face-to-face and video-recorded sessions and collect 371 individual behavior data which we were able to analyze and categorize. With the use of the app, our study contributes to the emerging body of evidence in categorizing children with PIMD/SMID's expressive behaviors which can be of great help in designing and planning interventions. Our findings revealed that children's needs were manifested mainly through hand and body movements and vocalization, which is similar, in partial, with the findings of the study done by Ashida and Ishikura (2013) [9]. This emphasizes the need to develop a system that predicts children's needs through speech or movement patterns. One of the recently developed technologies to capture human movements is the optical motion capture system [25], in which outputs can be analyzed using trajectory analyses, a powerful tool in motor behavior researches [26]. Outputs can be used in developing algorithms using a machine or deep learning methods. Further, to eliminate the digital divide, it is, therefore, necessary to not only predict the children's needs but also to develop a system that executes children's needs by transmitting data to Friendly VOCA, which voice outputs will be detected by smart speakers connected to devices and appliances. This will enable each child with PIMD/SMID to communicate and be mobile independently.

At present, we only rely on the interpretations of the children's close caregivers (e.g. parents, teachers, therapists, etc.) because the children are highly dependent on them for pervasive support in everyday tasks, 24 hours a day [1,4]. They are more capable to discern and interpret the mostly unique behavior of each child than other people. Thus, we are expecting that our system will help people who are not close to them to easily communicate with them and be part of their communication group.

### 4.1. Limitations

Despite the study's strengths, several limitations may affect the generalizability of our study findings. We were able to conduct multiple sessions among our target population, however, we only conducted them in a school setting. This limits our study in providing a more diverse perspective on children's behaviors as they have distinctive behaviors and needs at home and toward their immediate family members who they are more familiar with. This will be taken into consideration on our plans of testing the app at home and other locations as this will help us to measure the ability of the app in detecting and transmitting behavior, location, and environment data to the app database in a different setting. The children were recruited from a special needs education school which limits our findings to children with PIMD/SMID who are attending special needs schools. We assume that there are children who don't attend special schools or who are attending other healthcare facilities, thus, we will consider including them in our future interventions. Another limitation of our study was the method that we used to measure the accuracy rate of the app in detecting and transmitting location and environment data from iBeacon, GPS, ALPS sensors, and OpenWeatherMap API data sources. While it is ideal to measure the app's accuracy by comparing it with other apps that use similar location and environment data sensing technologies, to the best of our knowledge, no other app has been developed with the same goals and functions as the ChildSIDE app. Consequently, we had no other means of measuring this function other than counting the data transmitted and detected by the app to the database. Moreover, our findings conclusions on the movements involved in the expressive behaviors of the children are limited among the children recruited in our study and must be interpreted with caution. Lastly, we consider language limitations on the translation of our data from Japanese to English. Although the data were translated by a bilingual translator, we still consider that there were words in Japanese that did not have equivalent or was difficult to translate in English. This also leads to the limitations on the generalizability of our findings and conclusions as it may only represent the children with PIMD/SMID among the Japanese population which may not the same with that of other countries.

### 4.2. Conclusion

This study confirms that the use of the ChildSIDE app is an effective method in collecting children's expressive behaviors than the conventional paper-based collection methods, with a low level of user error. While the app had difficulty in detecting and transmitting short-distance indoor location sensor data from iBeacon, which we will address in our development process, it can provide GPS location information and comprehensive environmental data associated with each children's behavior with above 93% accuracy rates. This study also adds to the emerging body of evidence in the possibility of categorizing and interpreting children with PIMD/SMID's expressive behaviors, which, based form our findings, emphasizes the need to develop a system that uses motion capture system and analyze them using motion trajectory analyses and develop algorithms using a machine or deep learning to predict children's needs in the future.

**Acknowledgments**

This study was supported by the Ministry of Internal Affairs and Communications' Strategic Information and Communications R&D Promotion Programme (SCOPE) (MIC contract number: 181609003; research contract number 7; 2019507002; 2020507001) which had no role in the design, data collection and analysis, and writing of this manuscript. The authors would like to thank all the children and their caregivers who participated in this study.

**Author Contributions**

**Von Ralph Dane Marquez Herbuela**: Validation, Formal analysis, Writing-Original Draft and Review and Editing, Visualization **Tomonori Karita**: Conceptualization, Methodology, Funding acquisition, Project administration, Supervision, Writing-Review and Editing, Resources **Yoshiya Furukawa**: Methodology, Validation, Formal Analysis, Investigation, Data Curation, Writing-Review and Editing **Yoshinori Wada**: Validation, Formal analysis, Data Curation, Writing-Review and Editing **Shuichiro Senba, Eiko Onishi, Tatsuo Saeki:** Conceptualization, Software, Resources, Project administration, Writing-Review and Editing

**Conflict of Interest**
None declared.

**Supplementary Materials:**
Supplementary File S1: Behavior data collected by ChildSIDE app.